\documentclass[12pt]{iopart}

\usepackage{iopams}
\usepackage{amsfonts}
\begin{document}

\newfont{\gotha}{cmfrak}
\newfont{\gothlarge}{cmfrak scaled \magstep5}

\title{Renormalization group running cosmologies - from a scale setting to 
holographic dark energy}

\author{Branko Guberina \footnote{Invited plenary talk given at the 
{\it $2^{nd}$ International Conference on Quantum
Theories and Renormalization Group in Gravity and Cosmology}, (IRGAC 2006),
Barcelona, Catalonia, Spain, July 11-15, 2006.}}

\address{Theoretical Physics Division, Rudjer Bo\v skovi\'c Institute, P.O.B.
182, HR-10002 Zagreb, Croatia}
\ead{guberina@thphys.irb.hr}

\begin{abstract}

A scale-dependent cosmological constant $\Lambda$ and the 
Newton constant $G$ emerge naturally in quantum field theory 
in a curved space-time background leading
to renormalization group running cosmologies. A scale-setting procedure is 
discussed in these cosmological models and the interpretation of the scale is 
emphasized. This setup introduces dark energy without invoking quintessence-like fields and can be applied to a variety of problems. The scale-dependent 
$\Lambda$ and $G$ are also naturally incorporated into the generalized 
holographic dark energy model, and applied to different aspects of cosmology.

\end{abstract}


---------------------------------

\maketitle

\section{Introduction}

The resurrection of the cosmological
constant\cite{Einstein,Edding,Landau}, - a consequence of the 
evidence
for the acceleration of the universe driven by nongravitating (unclustered)
dark energy with negative preasure \cite{Perl}, precise measurement of the
cosmic microwave background\cite{Benett,spe} - has become one of the most futile
playgrounds for the broad spectrum of new investigations
\cite{Nobbenhuis,Dolgov}. 
On the one hand,
there is experimental evidence for a very tiny but positive cosmological
constant, which, theoretically can be studied using the powerful tools of
general relativity, and from quantum field theory to supersymmetry,
superstrings, and
branes\cite{Witten,WeinbergCC,LisaR,Polchinski,
Susskind1,Burgess}. 
On the other hand, there is,
{\it horribile dictu}, a flagrant discrepancy
of the 123 orders of magnitude between the theoretical result and the
experimental value. 
In this paper we discuss the basic underlying principles and/or ans{\" a}tze
in the effective quantum field theory on curved space-time, which is one of
the possible frameworks used to study the running of the cosmological constant.

We would like to answer the following questions:
Do we have a tool/framework which we can use in calculation of the 
cosmological constant properties 
without solving the $10^{123}$ discrepancy? Are we able to reconcile general 
relativity with quantum field theory (QFT) in spite of the fact that such 
a theory appears to be nonrenormalizable\cite{Einstein}?
Especially, can one build a reliable theory at low energies 
(large distances) which can unite an obviously successful general 
relativity with QFT?

\section{Cosmological constant renormalization and the decoupling theorem}

\begin{flushright}
$\mbox{\gotha { 
 Ich habe wieder etwas verbrochen  
in der Gravitationstheorie,}}$ \\ 
$\mbox{\gotha { was mich ein wenig in Gefahr bringt, 
 in ein Tollhaus interniert zu werden.}}$ \\

{\footnotesize A. Einstein, a letter\footnote{{\it I have again perpetrated
something relating to the theory of gravitation that might endanger me of
being commited to a madhouse}, quoted as in \cite{Straumann3}.}  to P. Ehrenfest, 1917}
\end{flushright}

By an 'effective field theory' (EFT) we understand a full quantum 
field theory with loops, regularization of divergences, 
renormalization, etc. It is basically 
the uncertainty principle that splits the theory in 
two regimes, so that EFT 'decouples' from the high energy
sector\cite{Donoghue}.
All effects of heavy particles appear in loops and the short-distance
physics is described by a local lagrangian which generally contains
 infinitely many local terms. The most general local lagrangian 
$\mbox{\gotha L}$ 
describes the high energy behavior of the theory\cite{WeinbergEFT}. 
But can we calculate anything with an infinite number of terms in 
$\mbox{\gotha L}$?

The second very important ingredient which resolves the problem of 
infinitely many terms in the theory is the fact that a local 
lagrangian is an energy expansion, the expansion parameter being the 
ratio of low energy scale and high energy scale. 
This reduces an infinite number of terms in lagrangian to the 
first few terms which can be used in calculation. In the quantized theory
there appear loops and ultraviolet divergences can be removed using the 
counterterms which are exactly of forms which are present in
the most general local lagrangian.

A very well-known example of an effective field theory used instead of
a complete renormalizable theory is the heavy quark effective theory
\cite{Wise}. 
The latter is a nonrenormalizable theory but exhibits the properties of 
heavy quarks explicitly in a transparent way, with the rest of physics 
left as a few parameters. In the end the theory proved to be a very
successful {\it quid pro quo}.

Another example which serves as a benchmark for EFT is the chiral 
perturbation theory (CPT)
\cite{ChPT}, which is a nonlinear realization of a 
low-energy limit of QCD. It is a complete QFT, calculated to one- and
two-loops and compared with experiment. It appears that some 
rare processes which are absent at the tree level have contributions
from the loops - this answers a sceptical questioning, namely the 
{\it raison d'\^{e}tre} of calculation of loops in a nonrenormalizable theory.

The effective field theory of gravity has been pushed up by Feynman 
\cite{Feynmann}, de Witt \cite{deWitt}, 't Hooft and Veltman 
\cite{tHooftVeltman},
etc., and was focused on high energy aspects of the theory. The 
program encountered serious difficulties with the divergence 
structure and did not lead to a satisfactory quantum gravity.

The low energy formulation of quantum field theory on curved 
space-time
\cite{BirellDavies}
\cite{Buchbinder}  starts with the Einstein-Hilbert action plus the 
gauged matter lagrangian. The theory is quantized in such a way 
that the background field method preserves the 
symmetries of general relativity, and still allows to 
gauge-fix  quantum fluctuations. 
By inspection of the results obtained for the 
graviton-graviton scattering, one encounters the non-analytic
terms in the logarithms - a clear signal of long-distance 
effects in quantum gravity\cite{Donoghue}.

The vacuum action necessary to ensure the renormalizability 
of the gauged scalar (matter) Lagrangian should contain terms:
$ R^2_{\mu\nu\rho\sigma}$, $R^2_{\mu\nu}$, $R^2$, and $\Box R$
\cite{ShapSolaPL}\cite{ShapSolaJHEP}, and a trace anomaly
term\cite{Anton,Mottola}. 

The vacuum action is then given as

 \begin{eqnarray}
 {\mathfrak A}_{vac}  =  \frac{1}{16 \pi G} \int d^4 x \sqrt{-g} \left[(R - 2\Lambda ) + 
    ( a_1 R_{\mu\nu\rho\sigma}^2 
   + a_2 R_{\mu\nu}^2 +a_3 R^2 + a_4 \Box R ) \right] \, \nonumber \\
    + {\mathfrak A}_{anom}.
 \end{eqnarray}
where ${\mbox{\gotha A}}_{anom}$ is a trace anomaly term.
Like the anomalous effective action in QCD, which appears as a consequence of
chiral anomaly in QCD, the term ${\mbox{\gotha A}}_{anom}$ should be included
in EFT of gravity coupled to matter, even in zero momentum limit. It is the
renormalization of the stress tensor that generates the trace anomaly, and
provides us with the possibly important effects at large distances. This is
true even if one had a theory with only massive particles - in that case the
fluctuations of metric would generate such a term.

All divergences can be removed by renormalization of the matter fields,
their masses and couplings, the bare parameters. The matter (scalar) 
action  is given by 

\begin{equation}
{\mathfrak A}_m = \int d^4x \sqrt {-g} {\mathfrak L}_m =
\int d^4x \sqrt {-g} [ \frac{1}{2} g^{\mu\nu} (\partial_{\mu}\phi
\partial_{\nu}\phi ) - V(\phi )] \, .
\end{equation}
$V(\phi )$ can be taken to be, e. g., the following expression:

\begin{equation}
V(\phi ) = -\frac{1}{2} m^2_0 \phi^2 - \frac{g_0}{24} \phi^4 - \bar{\Lambda}_0 
+ \eta_0 \phi ,
\end{equation}
where $\bar{\Lambda}_0 = (8 \pi G)^{-1} \Lambda_0$ is an arbitrary constant,
which is basically the same as the cosmological constant in the Einstein
lagrangian. Taking, for simplicity, $g_0 = 0$ (free scalar field) and
$\eta_0 = 0$, one can calculate the lowest-order vacuum energy. After 
the renormalization one obtains the {\it physical} vacuum energy
density given as\cite{Brown}

\begin{equation}
{\mathfrak E} = \bar\Lambda (\mu ) + \frac{m^4}{4(4\pi )^2} 
[\ln \frac{m^2}{\mu^2} - \frac{3}{2} ] .
\end{equation}
Here, the renormalized and bare CC are related as $
{\bar\Lambda}_{bare} = \mu^{d-4} (\bar\Lambda + z_\Lambda m^4)$,
where $z_\Lambda$ is a counterterm, $z_\Lambda = - \frac{1}{4(4\pi )^2} 
\frac{1}{\epsilon}$.
It is clear that the vacuum energy $\mbox{\gotha E}$, as given in (4) 
at the one-loop level, is independent 
of the arbitrary renormalization scale $\mu$. The $\mu$-dependence 
of $\Lambda (\mu )$ cancels the 
$\mu$-dependence of $\ln \frac{m^2}{\mu^2}$. 

It is now easy to derive the renormalization group equations for the CC.
Normally, one expects that heavy particles decouple in the theory, 
according to the Appelquist-Carazzone decoupling theorem \cite{Appel}.
However, an interesting result of nondecoupling of heavy particles is
found by Babic {\it et al.}\cite{Babic}. Assume there exist two particles,
a heavy one with mass $M$ and a light one with mass $m$. Then, for 
$m \ll \mu \ll M$ one expects the decoupling of a heavy 
particle with the suppression factor $\mu^2 /M^2$. Instead, one finds the
following behavior:

\begin{equation}
\label{eq:lightheavy}
(4\pi )^2 \mu \frac{\partial }{\partial \mu }\Lambda(\mu) =
\frac{1}{2}a \frac{\mu^2}{M^2} M^4 + \frac{1}{2} m^4  .   
\end{equation}
Obviously, the suppression factor $\mu^2 /M^2$ ($a$ is the number of order $\mathcal{O}(1)$) is not sufficient to
suppress the contribution of the heavy scalar, because
\begin{equation}
\mu^2 M^2 \gg m^4 .
\end{equation}
The net result of \cite{Babic,GuberinaFate} is that one actually has to take into account
the whole spectrum of heavy particles (which we do not know). 

The question intimately connected with the running parameters is about
 scale fixing. We may argue, {\it per analogiam} with QCD, that the scale
should be a typical momentum of the particles involved in a given 
physical process. In our case, this would be, for example, typical momenta
of the gravitons involved in the loop calculation. Actually, in QCD, 
the running 
scale is obtained by looking at the scaling properties of 
Green functions. All 
momenta are scaled according to $p_i \rightarrow \lambda p_i$, where
 $\lambda$ is a scale parameter. In QFT on curved space-time one is doing 
exactly the same. The transformation $g_{\mu\nu} \rightarrow \lambda^2 
g_{\mu\nu}$ implies $p^2 \equiv g_{\mu\nu} p^\mu p^\nu \rightarrow  
g_{\mu\nu} (\lambda p^\mu ) (\lambda p^\nu )$, in full analogy with
QCD.

However, there is a serious problem because there are no gravitons attached 
in the lowest order - therefore, no explicit momentum is determined. 
Even more, as shown by Gorbar and Shapiro \cite{Gorbar}, it appears impossible
to determine the lowest-order $\beta$-functions if calculation was performed 
on the flat background.  
\footnote{Gorbar-Shapiro calculation was performed in a physical
mass-dependent scheme and the behavior of the higher order terms in $\beta$
function in the infrared regime shows a clear decoupling - much the same
behavior which was predicted for $\Lambda /G$ by Babic et al. \cite{Babic} 
on intuitive basis. However, as pointed in \cite{Gorbar}, the absence of the
$\beta$-functions for $\Lambda /G$ and $G^{-1}$ is probably an artefact of the
perturbative expansion in $h_{\mu\nu}$, and not a fundamental property of the
RG in curved space-time.} 
One is therefore forced to rely on a certain intuitive educated guess
so as to determine the scale or to reinterpret the scale as an infrared 
cutoff, which is physically quite well founded and acceptable. We will
encounter and discuss this question in the next chapters, too.

\section{RG running cosmologies - a scale setting procedure}

\begin{flushright}

{\it Vos calculs sont corrects,
 mais votre physique est abominable.}\\
{\footnotesize A. Einstein to G. E. Lema\^{i}tre} \\

\end{flushright}

The RGE scale setting procedure is far from being obvious and
satisfactory. The question arises: is there a certain physical argument 
that would induce a procedure which might remove arbitrariness and lead 
to a scale setting?

The class of  RGE-based cosmological models have certain common 
properties. To simplify an argument, let us assume that there is only one 
universal running scale $\mu$, and the only running quantities are 
$\rho_{\Lambda} = \frac{\Lambda}{8 \pi G}$ and the Newton constant $G$. 
This means that we ignore, for example, the eventual mild dependence of 
the particle masses that appear in the theory, etc. A further assumption 
is that the ponderable matter and radiation evolve in a standard way; 
the energy-momentum exchange between these components and the dynamical
cosmological term is allowed. However, it is clear that one cannot talk 
about the conservation of energy and momentum for matter alone 
\cite{Einstein}. This follows from the vanishing of the covariant 
derivative of the mixed energy tensor of matter

\begin{equation}
\frac{\partial {\mathfrak T}_{\sigma}^{\mu}}{\partial x^{\mu}} 
- \Gamma^{\mu}_{\sigma\nu} {\mathfrak T}_{\mu}^{\nu}= 0 ,  
\end{equation}
where the energy-momentum density ${\mbox{\gotha T}^{\mu}_{\sigma}}$ is 
related to the energy-momentum tensor ${\cal T}^{\mu}_{\sigma}$ as 
${\mathfrak T}^{\tau}_{\sigma} = {\cal T}_{\mu\sigma} g^{\tau\mu}\sqrt{-g}$ .
By exerting forces upon 'matter' the gravitational field transfers energy 
to it - as is precisely described by the term $\Gamma^{\mu}_{\sigma\nu}
 \mbox{\gotha T}_{\mu}^{\nu}$ in (7).

The scale setting procedure will now be applied to two specific cosmologies:
the nonperturbative quantum gravity \cite{Reuter}, and the cosmological model
derived using quantum field theory on curved space-time 
\cite{ShapSolaJHEP}\cite{Babic}.

The RGE scale setting procedure is far from being obvious and 
satisfactory. The question arises: is there a certain physical argument 
that would induce a 
procedure which might remove certain arbitrariness and lead to a scale fixing?

Our input equation takes a form of the RGE improved Einstein equation

\begin{equation}
G_{\mu\nu} = - 8 \pi G(\mu) [T_{\mu\nu}^m + T_{\mu\nu}^{\Lambda} ] ,
\end{equation}
where 
\begin{equation}
G_{\mu\nu}  = R_{\mu\nu} - \frac{1}{2} g_{\mu\nu} R
\end{equation}
and $T_{\mu\nu}^{\Lambda}(\mu) = g_{\mu\nu} \rho_{\Lambda}(\mu)$ .
Here, $G_{\mu\nu}$ is the Einstein tensor, $R_{\mu\nu}$ and $R$ denote 
the Ricci
tensor and scalar, respectively, while $T_{\mu\nu}^m$ and 
$T_{\mu\nu}^{\Lambda}$ denote thematter and cosmological constant energy-momentum
tensors, respectively. To summarize, the only physical requirements to this 
equation
are: i. its general covariance, ii. the $\mu$-dependence of $G$ and
$T_{\mu\nu}^{\Lambda}$, and iii. the implicit time dependence of $\mu$.   
 
The conditions i. - iii. translate into

\begin{equation}
G(\mu)[\rho_m + \rho_{\Lambda}(\mu)] + G(\mu) \dot{\rho}_{\Lambda}(\mu) 
= 0,
\end{equation}
where dots denote time derivatives. Assuming the nonvanishing $\dot{\rho}$ 
throughout the evolution of the Universe (which seems to be a reasonable 
assumption) suggests the matter density equation
\cite{Babic2}

\begin{equation}
\rho_m = -\rho_{\Lambda}(\mu) - G(\mu) (\frac{d \rho_{\Lambda}(\mu)}
{d \mu}) (\frac{d G}{d \mu})^{-1} .
\end{equation}

The r.h.s. of the (11) is a function of the scale factor $a$ since we assume
the canonical behavior of $\rho_m$, i. e., $\rho_m = \rho_{m,0} 
(\frac{a}{a_0})^{-3(1+w)}$ .
The r.h.s. is, however, a function of $\mu$, i. e., (11) has a form
$\rho_m = f^{-1} (\rho_m)$ .

It is important to stress that our procedure lacks the first-principle 
connection to quantum gravity and, therefore, is not a fundamental one. 
However, as long as $\rho_m$ in (11) retains its canonical form, the scale
is univocally fixed\cite{Babic2}.

\subsection{Nonperturbative quantum gravity}

This theory is based on the exact renormalization group approach,
applied to quantum gravity \cite{Reuter}. The keystone of the theory is the
 effective average action $\Gamma_k [g_{\mu\nu} ]$ which is basically a 
 Wilsonian coarse-grained free energy \cite{Reuter,BonReut}. The momentum scale $k$ is
 then interpreted as an infrared cutoff - for a physical system with a size $L$,
 the parameter $k \propto 1/L$ defines an infrared cutoff.
 The path integral which defines the effective average action
 $\Gamma_k [g_{\mu\nu} ]$ integrates only the quantum fluctuations with the 
 momenta $p^2 \ll k^2$, thus describing the dynamics of the metric 
 averaged over the volume $(k^{-1})^3$.
The theory is valid near the scale $k$ in the sense that for any scale
$k$ there is a $\Gamma_k$ which is an effective field theory 
at that scale.

All gravitational phenomena are correctly described at tree level by 
$\Gamma_k$ including the contributions of loops with $p^2 \geq k^2$.
This means that all quantum fluctuations with $p^2 > k^2$ 
are integrated out. This is very similar to the effective QCD,
where high-energy quarks and gluons are integrated out.
The large-distance metric fluctuations, $p^2 \leq k^2$, are not
included as expected. However, in the limit $k\rightarrow 0$, the 
infrared cutoff disapears and one recovers the original action 
$\Gamma$.

From the physical point of view, the infrared cutoff in nonperturbative
quantum gravity corresponds physically to the dimension of the system.
Its determination is not trivial - in massless theories such as massless
QED, its interpretation is clear because $k^{-1}$ is the only mass scale 
present in the theory. In reality, a variety of mass scales is present
and caution is demanded.

The correct way to proceed is to study the RG flaw of the effective 
action $\Gamma_k [g_{\mu\nu} ]$ and identify the infrared cutoff
by inspecting the RG evolution. Once the infrared cutoff is fixed,
one should solve the Bianchi identities and the conservation laws
for matter \cite{ReuterWeyerJCAP12(2004)001}. 

One can also start {\it vice versa}: first, use the Bianchi identities
to fix a scale and, then, look for a meaningful physical interpretation
of the scale\cite{Babic2}.

In nonperturbative quantum gravity  the choice $k\propto 1/t$,
where $t$ is a cosmological time, 
i. e., the temporal distance between a given event
and the big bang, seems very plausible. In EFT one integrates 
quantum fluctuations with momenta smaller than $1/t$, since these
fluctuations should not play any role yet. This is a 
physical meaning of the infrared cutoff.

In the QG setting \cite{BonReut}, the infrared cutoff plays the role of 
the general RG scale $\mu$. The application of the scale-setting 
procedure results in expressing the scale $k$ in terms of 
cosmological parameters

\begin{equation}
k = ( 8\pi g_* \lambda_*^{-1} \rho_m )^{1/4}
\end{equation}
where $g_*$ and $\lambda_*$ are constants related to $G$ and $\Lambda$,
respectively, in the following way: $\Lambda (k) = 
\lambda_* k^2 , \,\,\,\,  G(k) = g_* k^{-2}$.
The values of $k$ for different scenarios are obtained in
\cite{Babic2} and compared with the results of 
\cite{BonReut}.

For $\Omega_K = 0$, and arbitrary $w$, our result for $k$ agrees with the 
result of \cite{BonReut}. The same happens to be true for $w=1/3$ and 
arbitrary $\Omega_K$. However, for $w=0$ and $\Omega_K \geq 0$, 
i. e., for the universe containing nonrelativistic matter only, and 
having arbitrary positive curvature, one obtains the law of evolution of 
the scale factor as a function of time.

This result is somewhat different from the result of \cite{BonReut}.
In their calculation consistent solutions with $K=+1$ or $K=-1$ exist
only for a radiation-dominated universe. On the contrary, our
procedure, which is obviously mathematically correct, leads to 
consistent solutions for a universe having arbitrary curvature and 
for arbitrary $w$.

Is it possible that our solutions, although mathematically correct, to 
paraphrase Einstein, are {\it abominable}?

The answer is largely discussed in \cite{Babic2}, and to shorten the argument, 
we emphasize that the scale-setting procedure always leads to
a mathematically correct consistent fixing of $k$. However, at the
same time, the physically acceptable choices for the scale $k$ are 
only those  having a geometrical interpretation. As a matter of fact, 
we argued in \cite{Babic2} that for the universe with small
curvature, the scale $k$ is reasonably close (in a regime with the IR 
fixed point domination) to the scale obtained for a flat universe
and, therefore, a satisfactory geometrical interpretation is 
still possible.

\subsection{Models from quantum field theory on curved space-time}

Let us consider  a generic case when both $\rho$ and $\Lambda$ 
can be expanded in series in $\mu^2$, where $\mu$ is the RGE scale.
These models are studied in the framework of QFT on curved space-time,
with a correct treatment of heavy-particle decoupling.

When the scale $\mu$ is smaller than all masses in the theory, the 
coefficients in the expansion, $C_i$ and $D_i$ can be either 
calculated or estimated. The expansion is given by

\begin{equation}
\label{eq:Lamexp}
\rho_{\Lambda} = \sum_{n=0}^{\infty} C_{n} \mu^{2n} \, , \, \, \, \, 
G^{-1} = \sum_{n=0}^{\infty} D_{n} \mu^{2n} \, .
\end{equation} 

The application of the scale-setting procedure \cite{Babic2} yields the
identificationof the scale $\mu$
\begin{equation}
\label{eq:rhommu}
\rho_{m} = -C_{0} + \frac{C_{1} D_{0}}{D_{1}} + 2 \frac{D_{0}}{D_{1}}\left(
C_{2} - \frac{C_{1} D_{2}}{D_{1}} \right) \mu^{2} + \dots \, , 
\end{equation}

Generally, in these cosmologies, $C_1 \sim m^2_{max}$, 
$C_2 \sim N_b - N_f \sim 1$, $C_3 \sim 1/m^2_{min}$,
etc. Here $m_{max}$ and $m_{min}$ are the largest and the smallest
mass, respectively.

The same reasoning applies to $G$.
The qualitative analysis of (14) shows \cite{Babic2} that the 
results obtained are at variance with observational bounds. By the
'qualitative analysis' we mean the fact that the coefficients 
$C_i$ and $D_i$ are determined on dimensional grounds only.

\section{Holographic Dark Energy}

The holographic principle is based on the assumption that QFT 
overcounts the true degrees of freedom - therefore some extra 
nonlocal constraints are necessary to obtain a reliable 
effective field theory.

The entropy S scales extensively in an effective quantum field
theory: for a system of size $L$ with the UV cutoff $\Lambda_{UV}$,

\begin{equation}
S \sim L^3 \Lambda_{UV}^3 . 
\end{equation}

However, it is known that, according to Beckenstein \cite{Beck},
the maximal entropy in a box of volume $L^3$ grows only as the area 
$A$ of the box. This means that, for example, all information that can be
present in a black hole, should be coded on 
the two-dimensional horizon (surface) 
in Planckian pixels \cite{Suss} 
\footnote{The finitness of the Universe (the finite age and the finite
particle horizon) leads to the upper limit in information (a number of bits)
inside the horizon volume - for the universe at present time it amounts to
approximately $10^{123}$ \cite{Lloyd}. This limit which is basically a
consequence of the holographic principle, can have a more profound
implications for fundamental physics, as discussed recently by
Paul Davies in \cite{Davies2007}, see also the references therein.}. 
The underlying principle was dubbed a 
holographic principle - a connection to a holographic display of 
'a very pretty flirtatious girl' in the Stanford Physics Department
was a rather recent revelation \cite{Suss}. A connection to physics
was clear previously - a fact that measuring everything on the surface 
with Planckian resolution, one can reconstruct everything inside the 
volume - pointed to a very well known notion of hologram. 
The idea of holography is also intimately related to  
the ultraviolet/infrared connection \cite{SussW}, i. e., 
to the fact that going to higher and higher UV energies (short 
distances) one actually probes the long-distance physics -
 at very high 
energies, black holes would be created, which would emit 
long-wavelength quanta. 

On the other hand, it was known \cite{HooftSuss} that $3+1$ QFT's 
overcome the the degrees of freedom, and a local QFT cannot describe 
quantum gravity, because it has too many degrees of freedom in UV.

The exit, suggested bu Cohen {\it et al.} \cite{Cohen}, is
to limit the volume of the system according to

\begin{equation}
L^3 \Lambda^3 \leq S_{BH} \equiv \pi L^2 M_P^2 ,
\end{equation}
where $L$ is the size of the box, and $S_{BH}$ is the entropy 
of the black hole. 
Obviously, the black-hole entropy grows as area ($\sim L^2$) of
the horizon surrounding it
\footnote{Assuming the dominant energy condition, $\rho + p \geq 0$, 
Davies\cite{Davies88} showed that the cosmological event horizon area 
of the Friedmann-Lema\^{i}tre universe never decreases, {\it per analogiam} 
with Hawking's area theorem for black holes\cite{Hawking72}. Even more, 
the cosmological event horizon increases also in models in which the radius 
of event horizon decreases. Actually, the loss of entropy from within the 
cosmological horizon (due to the matter, radiation and/or black holes 
crossing the cosmological horizon) is compensated by an increase in
cosmological event horizon entropy - quite in agreement with the
generalized second law of thermodynamics\cite{Davies2003},
\cite{Padman2002},\cite{Susskind95}}.

If inside the volume $V \sim L^3$ one were able to find a region 
with an entropy larger than the entropy of a black hole of the
same size $L^3$, but with smaller energy than $E_{BH}$, 
this would immediately lead to  a violation of the second law of 
thermodynamics. Namely, by adding an additional matter to a box
$L^3$ one would eventually form a black hole, but with a smaller 
entropy than the original entropy. 

Cohen {\it et al.}\cite{Cohen} realized that the constraint (16) 
unavoidably included many states with the Schwarzschild radius
larger than the box size $L$. Therefore, the introduction of an 
infrared cutoff $1/L$ which {\it excludes} all states  that lie
within their Schwarzschild radius was a necessity. A new constraint
reads

\begin{equation}
L^3 \Lambda^4_{UV} \leq L M^2_P ,
\end{equation}
i. e., the entropy in a given volume $L^3$ should not exceed
the energy of a black hole of the same size $L$. An immediate 
consequence is that the IR cutoff now scales as $\Lambda^{-2}_{UV}$
\cite{Cohen}. To summarize: It is obvious that the usual 
quantum corrections to the vacuum energy density (zero-point energy)
give a wrong prediction. A holographic principle has, however,
an intuitive physical idea behind it - the idea that the fields at
the present energy scales do not fluctuate independently over the entire
horizon, or even over the universe - the idea we have already 
encountered in the basis of some theories, such as nonperturbative 
quantum gravity \cite{BonReut}. Actually, if one takes the infrared 
cutoff to be of the order of $H^{-1}$, i. e., approximately 
the size of the present horizon, the value of $\Lambda^4_{UV}$ 
is coming down to something of the order of $(meV)^4$ - the 
right size of the present cosmological constant.

A number of authors have developed what is called a generalized 
holographic dark energy\cite{DE} - a model where both the cosmological 
constant (CC) and $G$ within QFT on curved space 
are runing. Applying the 
relation of Cohen {\it et al.} between the UV and IR cutoff 
results in an upper bound to the zero-point energy density $\rho_\Lambda$,
given as 

\begin{equation}
\rho_\Lambda \leq \mu^2 G^{-1}_N (\mu) .
\end{equation}
This is a generalized dark-energy model, where the RG scale $\mu$ 
is promoted to the IR cutoff \cite{Horvat,GuberinaJCAP}. 
In this approach, $\mu = 1/L$ is taken 
to be an IR cutoff and, therefore, the interpretation of 
the scale $\mu$ is restricted to this definition, Again, taking 
$L = H_0^{-1} = 10^{28} cm$ leads to the present observed value for 
the dark-energy density $\rho_\Lambda = 10^{-47} GeV^4$.

The precise choice of $\mu$ is a sensitive question. If $\rho_\Lambda$
is considered to be the energy density of a noninteracting perfect
fluid - then the choice $\mu = L^{-1}$ fails to recover the 
equation of state (EOS) for a dark-energy dominated universe, cf. 
the work of Hsu \cite{Hsu}. 
Even more, choosing $L = H^{-1}$  always leads to 
$\rho_\Lambda = \rho_m$ for flat space, thus hindering a 
decelerating era of the universe for redshifts $z > 0.5$. 
However, a correct EOS is obtained if one chooses a future event 
horizon for an infrared cutoff, as pointed by Li\cite{Li}.

Since (18) is derived using the ZPE, the natural interpretation
of dark energy in this equation is through the variable, or
interacting CC with $w=-1$. 

The energy transfer between various components in the universe 
(including a case with $G$ varying with time) is described by 
a generalized equation of continuity

\begin{equation}
{\dot G}_N (\rho_\Lambda + \rho_m ) 
+ G_N {\dot\rho}_\Lambda + G_N ({\dot\rho}_m + 3H\rho_m ) =0 .
\end{equation}
Here one should notice that $\rho_\Lambda$ is affected not only
by matter, but also by a time-dependent $G_N$. In addition, (19) 
leads to the conservation of the quantity $G_N T_{\mu\nu}^{total}$. 
\footnote{Sourced Friedmann eqs. with holographic dark energy are
studied by Myung \cite{Myung}}.

The holographic restriction (18) and the generalized equation of 
continuity (19) were used in \cite{Babic2} in order to constrain 
the parameters of of the RG evolution in QFT in curved-space background.
It was asumed that the scale dependence of the CC and $G$ arose 
solely from particle field fluctuations, i. e., no quintessence-type
scalar fields were present.
Again, one assumes the usual RG laws for the RG running scale $\mu$
below the lowest mass in the theory. It is important to note that the
scale $\mu$ cannot be set from the first principles. Estimating the 
coefficients $C_i$ and $D_i$ on dimensional grounds, and noting that 
$C_0$, i. e., the vacuum ground state of the CC, 
coincides here with the IR limit of the CC, one is able to
give a qualitative analysis. The context is set by fixing the 
matter density law to be a canonical one, i. e., no energy transfer 
between matter and  other components is allowed. This further reduces
(19), which, after insertion of the holographic expression (18), 
leads to the scale $\mu$ given as

\begin{equation}
\mu = - \frac{1}{2} G'_N (\mu ) \rho_m .
\end{equation}

Equation (20) shows that once $G_N (\mu )$ is known, the IR cutoff
$\mu$ is fixed. For $\mu > 0, G'_N > 0$, then $d/dt G_N > 0$, i. e.,
$G_N (t)$ increases with increasing cosmic time $t$. This implies
that $G_N$ is asymptotically free - the property seen in quantum 
gravity models at the one-loop level, cf. \cite{Fradkin} This is 
an interesting phenomenon as the asymptotic freedom is of some 
interest for galaxy dynamics and rotation curves \cite{ShaSoStef}.

A more interesting case is the one with variable $G_N$ and $\rho_\Lambda$,
 and the canonical law for matter. Inserting the expansions
in $\mu$ for $\rho_\Lambda$ and $G_N$ into the scaling fixing relation 
(25) leads to the expression for the scale $\mu$. Using the
estimates for $D_n$, one obtains
\begin{equation}
\mu^2 \approx \frac{1}{2} m^2_{min} ( 1 -  M^{-1}_{Pl} \rho_m ).
\end{equation}
The following remarks are in order. The value of $\mu$ is marginally
acceptable as far as the convergence of the $\rho_\Lambda (\mu )$ 
and $G_N (\mu )$ series is concerned. In addition, since the first 
time derivative of $G_N$ is negative, one obtains $D_1 \approx C_2 
> 0$. Furthermore, (26) shows a very slow variation of the scale
$\mu$ with the scale factor $a$ or the cosmic time $t$. 

However, once the RG scale crosses below the smallest mass in 
the theory, it effectively freezes at $m_{min} \sim H_0 \sim 10^{-33}
eV$. This is the main result of holography\footnote{Further developments along
these lines are given in \cite{Nikolic}.}
 - one finds a hint for possible
quintessence-like particles in the spectrum. What holography basically does - 
it expands the particle spectrum to the extremum - the 
largest possible particle masses approach the Planck mass, and 
the smallest possible particle masses are around the lowest 
mass scale in the universe, $m_{min} \sim H_0$. The present value of 
the vacuum energy density appears as the product of squared masses
of the particles lying on the opposite sides of the spectrum - a hint to a
natural solution to the coincidence problem! 
\footnote{In a different context, Pavon and Zimdahl \cite{Pavon} showed that 
interaction between dark matter and dark energy could also lead 
to the solution of the coincidence problem.}

\section*{Acknowledgments}
I would like to thank  the organizers of the
IRGAC 2006 Conference for invitation. I am greatly indebted to 
Ana Babi\'c, Raul Horvat, Hrvoje Nikoli\'c, and Hrvoje \v Stefan\v ci\'c
 for numerous discussions and pleasant collaboration.
This work was supported by the Ministry of Science,
Education and Sport
of the Republic of Croatia under contract No. 098-0982930-2864, and
partially supported through the Agreement between the Astrophysical
Sector, S.I.S.S.A., and the Particle Physics and Cosmology Group, RBI.

\end{document}